\documentclass[]{aa} % for a non-referee version
\usepackage{graphicx}
\usepackage{natbib}
\usepackage{txfonts}
\newcommand{\PLowMADFive}{3.15}
\newcommand{\PHighMADFive}{3.4}
\newcommand{\PLowMADFour}{3.4}
\newcommand{\PHighMADFour}{3.7}
\newcommand{\PLowMADFourFive}{3.26}
\newcommand{\PHighMADFourFive}{3.49}

\newcommand{\PlanetSynch}{0.0015}

\begin{document}
\title{{\it MOST}\thanks{Based on data from the {\it MOST} satellite, a Canadian Space Agency mission jointly operated by Dynacon, Inc., the University of Toronto Institute of Aerospace Studies, and the University of British Columbia, with assistance from the University of Vienna, Austria.} detects variability on $\tau$ Bootis possibly induced by its planetary companion}

\author{ Gordon A.H. Walker\inst{1}, Bryce Croll\inst{2}, Jaymie M. Matthews\inst{3}, Rainer Kuschnig\inst{4}, Daniel Huber\inst{4},   Werner W. Weiss\inst{4}, Evgenya Shkolnik\inst{5}, Slavek M. Rucinski\inst{6}, David B. Guenther\inst{7}, Anthony F.J. Moffat\inst{8}, Dimitar Sasselov\inst{9}}

   \offprints{gordonwa@uvic.ca}
   \institute{1234 Hewlett Place, Victoria, BC V8S 4P7, Canada              \and
Dept. of Astronomy \& Astrophysics, Univ. Toronto, 
50 George St., Toronto, ON M5S 3H4, Canada
\and
Dept. of Physics \& Astronomy, UBC, 
6224 Agricultural Road, Vancouver, BC V6T 1Z1, Canada
\and
Institut f\"ur Astronomie, Universit\"at Wien 
T\"urkenschanzstrasse 17, A--1180 Wien, Austria
\and
NASA Astrobiology Institute, University of Hawaii, Manoa
\and
Dept. of Astronomy \& Astrophysics, David Dunlap Obs., Univ. Toronto 
P.O.~Box 360, Richmond Hill, ON L4C 4Y6, Canada
\and
Department of Astronomy and Physics, St. Mary's University
Halifax, NS B3H 3C3, Canada
\and
D\'ept de physique, Univ de Montr\'eal 
C.P.\ 6128, Succ.\ Centre-Ville, Montr\'eal, QC H3C 3J7, and Obs du mont M\'egantic, Canada
\and
Harvard-Smithsonian Center for Astrophysics, 
60 Garden Street, Cambridge, MA 02138, USA
             }

\abstract
{There is considerable interest in the possible interaction between parent stars and giant planetary companions in 51 Peg-type systems.}
{We shall demonstrate from $MOST$ satellite photometry and Ca II K line emission that there has been a persistent, variable 'region  on the surface of $\tau$ Boo A which tracked its giant planetary companion for some 440 planetary revolutions and lies $\sim$68$^{\circ}$ ($\phi=0.8$) in advance of the sub-planetary point.}
{The light curves are folded on a range of periods centered on the planetary orbital period and phase dependent variability is quantified by Fourier methods and by the mean absolute deviation ({\it MAD}) of the folded data for both the photometry and the Ca II K line reversals.}
{The  'region'  varies in brightness on the time scale of a rotation by $\sim$ 1 mmag. In 2004 it resembled a dark spot  of variable depth, while in 2005 it varied between bright and dark. The 2004 light curve gives a spot rotation period of 3.5 $\pm$0.7 d compared to the known planetary orbital period of 3.3125 d. The amplitude spectrum of the 2005 light curve shows no marked peak at the orbital period but the mean absolute deviation ({\it MAD}) of the light curve has a well defined maximum (half width $\sim$0.15 d) centered on the orbital frequency. Over the 123 planetary orbits spanned by the photometry the variable region detected in 2004 and in 2005 are synchronised to the planetary orbital period within 0.0015 d. The Ca II K line in 2001, 2002 and 2003 also shows enhanced K-line variability centered on $\phi=0.8$, extending coverage to some 440 planetary revolutions.}
{The apparently constant rotation period of the variable region  and its rapid variation make an explanation in terms of conventional star spots unlikely. The lack of complementary variability at $\phi$=$0.3$ and the detection of the variable region so far in advance of the sub-planetary point excludes tidal excitation, but the combined photometric and Ca II K line reversal results make a good case for an active region induced magnetically on the surface of $\tau$ Boo A by its planetary companion. }

   \date{Received ; accepted }


\authorrunning{Walker et al.}
\titlerunning{Variability on $\tau$ Bootis possibly induced by its planetary companion}
\maketitle

\section{Introduction}
\label{intro}

The detection of a giant planet extremely close (0.052 AU) to 51 Peg (\citealt{May97}; \citealt{Mar97})
anticipated the discovery of many more hot Jupiters near solar-type stars (see for example, \citealt{Mar00,But06}).
While some $\sim$20\% of the known extra-solar planets are in fact 51 Peg-like systems, this high percentage is at least partially due to an observational 
bias that favours detection of the large reflex velocities induced by close companions, rather than to such a truly high incidence. There is recent evidence, none the less, supporting a hot-Jupiter excess \citep{Gau06,But06} even after
the observational bias is taken into account.  A number of these close companions are seen in transit  \citep{Cha07} and the infrared spectral signatures of two of them have been detected directly, HD 209458b \citep{Ric07} and HD 189733b \citep{Gri07}, by the Spitzer Space Telescope.

$\tau$ Boo (HR 5185, HD 120136, F7 V, $V$=4.5) was announced in 1997 by \citet{But97} as a 51 Peg system with the large radial velocity semi-amplitude of 469 m s$^{-1}$, and  an orbital radius of 0.0462 AU corresponding to a minimum companion mass of 3.87 $M_{Jup}$ (more recently revised to 4.4 $M_{Jup}$  and a period of 3.3125 d by Butler in \citealt{Lei03}). \citet{But97} also noted that $\tau$ Boo rotates quite rapidly for a solar-type star ($v$sin$i$ $\sim$14 km s$^{-1}$) with a period \citep{Bal97} commensurate with the orbital period of the planet, suggesting tidal locking. 

Following the announcement by \citet{But97}, \citet{Hen00} carried out intensive Str\"omgren $b$ and $y$ photometry of $\tau$ Boo to set a limit to any transit. When they supplemented this with photometry from the six years 1993 to 1998, they found that the mean magnitude varied by $\sim$2 mmag from year to year and the semi-amplitude of any variation (including transits) synchronised to the planetary orbital period was 110 $\pm$90 micro mag. In addition, they had monitored Ca II H \& K using a narrow band filter on the Mount Wilson 100, and 60, inch telescopes for 13 years as part of the HK Project \citep{Bal98}. Apart from an unexplained $\sim$116 d period not seen in either photometry or radial velocities, they determined a rotation period of 3.2 d which they graded as `probable'. But the amplitude of the latter signal was small and they found the period ranged between 2.6 and 4.1 d which they felt might be the result of differential rotation. 

Later, \citet{Cun00} and \citet{Rub00} speculated that magnetic interaction between a giant planet and its primary in a 51 Peg system could generate detectable chromospheric activity and other effects such as those seen in HD 192263 by \citet{Hen02} and \citet{San03}. The interaction can be tidal and/or magnetic. If such planet-induced heating of the star lay in a narrow range of stellar longitude it would track the planet. This implies that tidally induced activity would have a period of $\sim$P$_{orb/2}$ while magnetic activity would have a period of P$_{orb}$. 

\citet{Shk03} and \citet{Shk05} specifically looked for signs of magnetic interaction in several 51 Peg primaries by monitoring Ca II H \& K reversals in high resolution spectra. They found two cases where chromospheric activity was apparently synchronised to the planetary orbital periods: HD 179949 (M$_{pl}$~sin$i$ = 0.98 M$_{J}$, P$_{orb}$ = 3.09 d, $a$ = 0.045 au) and $\upsilon$ And (M$_{pl}~$sin$i$ = 0.71 M$_{J}$, P$_{orb}$ = 4.62 d, $a$ = 0.059 au). \citet{Gu05} suggested that Alfv\'en waves generated by the planet's motion through the stellar magnetic field caused an  excitation focussed at the stellar surface. Unlike tidally induced activity which is expected at the sub-planetary point (where the line joining the centers of the star and planet intersect the stellar surface) and 180$^{\circ}$ from it, the longitude of magnetically induced activity would be dictated by the stellar magnetic field geometry. 

Despite the large mass of the planetary companion, \citet{Shk05}, like \citet{Hen00}, also found only a low level of chromospheric activity for $\tau$ Boo in 2001, 2002, and 2003 and they could see none obviously synchronised to the planetary period. They felt that the $\tau$ Boo system was an exception that proved the Alfv\'en wave model because tidal locking would avoid motion of the planet relative to the permanent component of the stellar magnetic field. 

>From spectropolarimetric observations with ESPaDOnS on the Canada-France-Hawaii Telescope, \citet{Cat07} recently detected a magnetic field of similar magnitude to the global solar value of a few gauss on $\tau$ Boo but with a more complex topology. Although their phase coverage was limited, they found that the Stokes $V$ profile variations were best modeled with a 3.1 d equatorial period, a 3.7 d polar period, and an inclination of the rotation axis to the line of sight $i$=40$^{\circ}$. They pointed out that this range of rotation periods is similar to that found by \citet{Hen00} from their long term monitoring of Ca II H\&K and clearly supports the suggestion of differential rotation.

We \citep{Wal06} reported an active photometric region on $\tau$ Boo synchronised to the orbital period of the planet based on photometric observations with the {\it Microvariability and Oscillations of STars} ({\it MOST}) satellite in 2004 and 2005.
Subsequently, a numerical error was found to have led to an inaccurate plot on which this conclusion was based. Here, the error has been corrected and we provide
further details of the observations and a more in-depth analysis from which we present an argument that there is indeed an active region on the star that rotates synchronously with the planet, some 68$^{\circ}$ in advance of the sub-planetary point. It is interesting to note that \citet{Shk05} found a similar positive advance in phase ($\phi=0.8$) for the active spot on HD 179949 induced by its planet. \citet{Cat07} speculate that, given their strong evidence for differential
rotation of $\tau$ Boo, the rotation period at the stellar latitude of the active region found by {\it MOST} must correspond to synchronism with the planet orbit at an intermediate latitude.

$\tau$ Boo was originally added to the {\it MOST} observing list as a priority candidate for the detection of phased, reflected signals from the planet \citep{Gre03}. It was also an interesting candidate in the search for solar-type p-modes, the primary {\it MOST} science goal. These will be the topics of later papers.

%______________________________________________ 
   \begin{figure*}[t]
   \centering
\includegraphics[height=0.8\textwidth, angle=270]{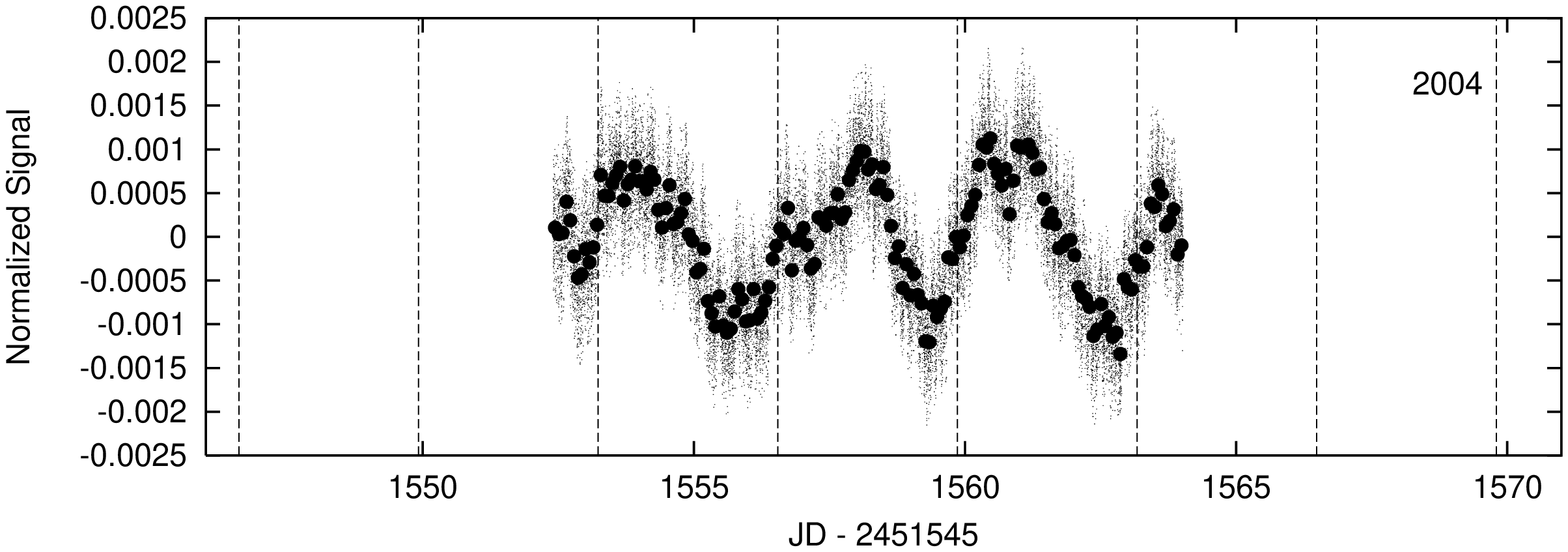}
\includegraphics[height=0.8\textwidth, angle=270]{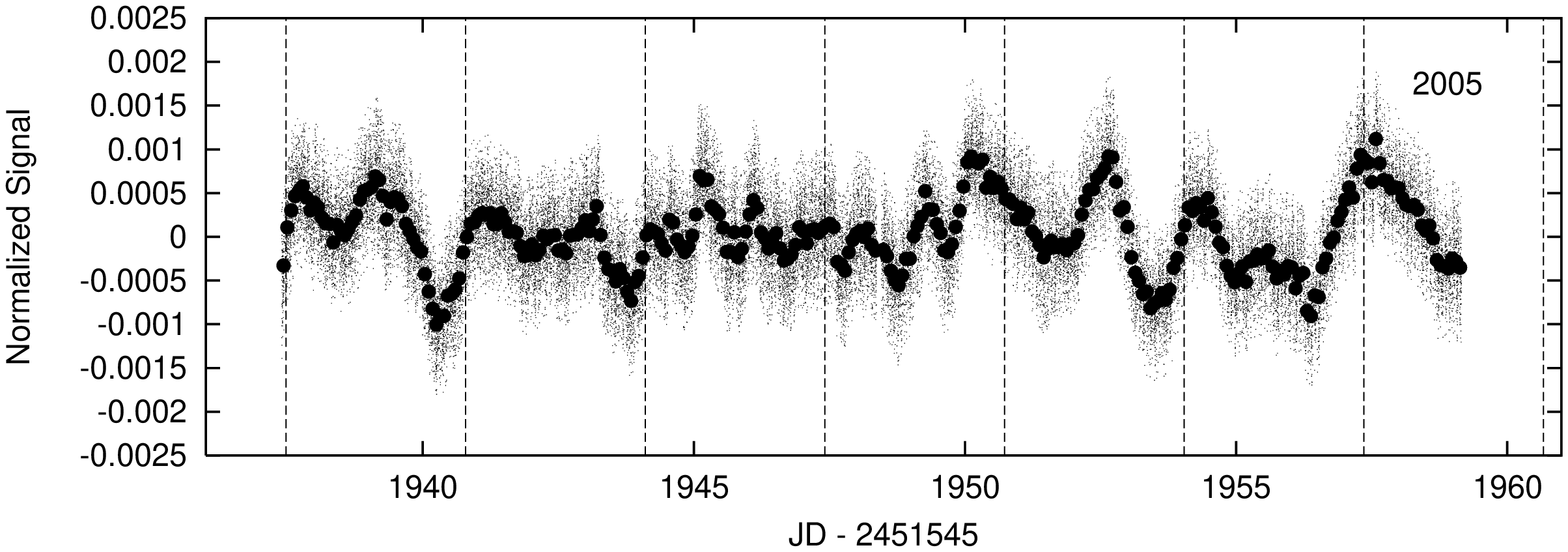}
\caption{{\it MOST} light curves for $\tau$ Boo from 2004 and 2005. The solid points are means from each 101.413 min {\it MOST} orbit.
	Individual observations are shown in grey. More details in Table 1 and text. The dashed vertical lines indicate zero phase (planet between observer and primary) of successive planetary orbits ($P$ = 3.3125 d).}
\label{Fig1}
\end{figure*}
%______________________________________________ 

\section{ {\it MOST} photometry and light curves}
\label{most}

The {\it MOST} satellite, launched on 30 June 2003, has been fully described
by \citet{MOST}. For the 2004 and 2005 observations of $\tau$ Boo, the 15/17.3 cm Rumak-Maksutov telescope fed two CCDs,
one for tracking and the other for science, through a single, custom, broadband filter (350 -- 700 nm) which has no analog in any other
photometric system. A Fabry image of the telescope entrance pupil covering some 1500 pixels  was projected onto the science CCD.
Since the experiment was designed to detect photometric variations with periods of minutes at micro-magnitude precision it does not
use comparison stars or flat-fielding for calibration. A dramatic reduction in tracking jitter early in 2004 to $\sim$1 arcsec led
to a significant improvement in photometric precision.

$\tau$ Boo A has an M2 dwarf stellar companion, $\tau$ Boo B, which is 4.2 mag fainter than, and 2.87 asec from, A \citep{Pat02}. B would always have been included with A in the 54 arcsec diameter photometric field stop and, even if variable, cannot have contributed significantly to the variations we detected in the light curve because of its faintness.

The observations reported here were reduced independently by RK (UBC) and DH (Vienna). In both cases, outlying data points
caused by poor tracking or cosmic ray hits were removed. At certain orbital phases, {\it MOST} suffers from parasitic light,
mostly Earth shine, the amount and phase depending on the stellar coordinates, spacecraft roll and season of the year.
To track the background, data are recorded for seven Fabry images adjacent to the target. In the UBC reduction \citep{Wal07},
 background signals are combined in a mean and subtracted from the target photometry - a procedure which simultaneously corrects for
bias, dark and background signals. 
%______________________________________________ Table OBS LOG
\begin{table*}[t]
\begin{center}
\caption{{\it MOST} Observations of $\tau$ Bootis
\label{TableObs}}

\begin{tabular}{cccccccc}
\hline
\hline
\noalign{\smallskip}
year	&dates &total &duty cycle	&exp &cycle&$\sigma^{a}$  &$\sigma^{b}$\\
	&HJD -- 2451545  &days &\% &sec&sec&mmag & mmag\\
\noalign{\smallskip}
\hline
\noalign{\smallskip}
2004 &1552.4 to 1563.9 &11.5 & 90&  25& 30& 0.12 &0.40 \\
2005 & 1937.4 to 1959.0 & 21.6 & 90& 25&30& 0.10 &0.35  \\
\noalign{\smallskip}
\hline
\noalign{\smallskip}
\multicolumn{8}{l}{$^{a}$ rms errors for data binned at the {\it MOST} 101.413 min orbital period.}\\
\multicolumn{8}{l}{$^{b}$ rms errors for the unbinned data.}\\

\end{tabular}  
\end{center}
\end{table*}

The Vienna reduction has been outlined in complete detail by \citet{Reg06}. It uses linear correlations of target and background
pixel intensities with the development of a step-wise multiple linear regression to remove only those target variations which also
occur in the background. The regression analysis provides the correct scaling to account for variations in the contamination level
from pixel to pixel. The multivariate solution has proved very effective in reducing the stray light contamination.

The $MOST$ light curves from both the UBC and Vienna reductions gave essentially identical light curves for both 2004 and 2005.
There was a brief observing run in 2006 covering less than two planetary revolutions. 
Early in 2006 the tracking CCD system of {\it MOST} failed due to a 
particle hit. Since then, both science and tracking have been carried out with the 
Science CCD system.
The 2006 observations of $\tau$ Boo were among
the first in which the single Science CCD was used simultaneously for tracking and science.
The 2006 data-set included a half day gap when time began
to be shared by chopping to an alternate field. During the half day gap the electronic board changed temperature and, because
of the long settling time, caused the photometric rms errors to double compared with 2004 and 2005.
Consequently, the 2006 data have not been included in this paper.

The light curves discussed in the rest of this paper and displayed in Figure 1 are all based on the Vienna reduction as we feel that it provides the more appropriate background elimination.

\subsection{Visual inspection and Fourier Analysis}
\label{SecFT}

A log of the {\it MOST} observations of $\tau$ Boo in 2004 and 2005 is given in Table 1. The 2004 run covered nearly
4 planetary orbits while that in 2005 covers nearly 7.
In neither set were there any gaps $>$ 3 hours. The reduced duty cycle is largely due to rejection of outliers during
South Atlantic Anomaly (SAA) passages and high stray light phases. For the analyses in this paper, data were binned at
the {\it MOST} orbital period of 101.413 min. Figure 1 displays full light curves for each year where time is given as
JD--2451545 (heliocentric). The orbitally binned data are the solid circles while the individual observations are faint, grey points. The vertical dashed lines correspond to zero phase (when the planet lies directly between the observer and the primary) of the planetary orbital period (3.3125 d).
The complete light curves can be downloaded from the MOST Public Archive at www.astro.ubc.ca/MOST.
The formal rms point-to-point precision in the individual observations, and the satellite orbital means, $\sigma$, are also
listed in Table 1.

\begin{figure}[htb]
   \centering
\includegraphics[width=0.5\textwidth, bb=12 150 560 610]{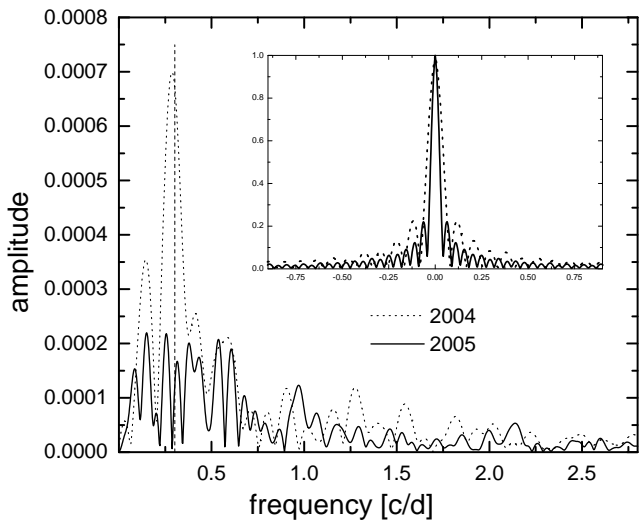}
\caption{Amplitude spectra for the {\it MOST} $\tau$ Boo light curves from each of 2004 (dots) and 2005 (solid). The vertical dashed line corresponds to the frequency of the planetary orbit, 0.302 c d$^{-1}$. The two window functions are shown in the insert.}
\label{FigFT}
\end{figure}

 The $\tau$ Boo light curve observed by {\it MOST} evolves rapidly
and changes in
shape in as little as one stellar rotation.
The light curves show peak to peak excursions $\sim$0.002 mag -
something difficult to track from a ground based observatory - but typical of the year to year
changes seen by \citet{Hen00}. In 2004 there is a clear set of four minima of variable depth
which occur at roughly the same planetary orbital phase. In the longer string of observations
in 2005, there are three or possibly four minima which resemble those in 2004 and occur at the
same planetary orbital phase. For the remaining two or three planetary revolutions there
is either a maximum or no obvious deviation at that phase.  Figure \ref{FigFT} shows the
Fourier spectra for both light curves together with their respective window functions.
The window function is the Fourier spectrum in which every data point is set to 1 thereby
providing a
measure of the frequency resolution. A linear trend was removed from the 2005 data before
performing the transform.  The vertical dashed line indicates the planetary orbital period
 of 0.302 cycle d$^{-1}$ (3.3125  d). The amplitude spectrum for 2004 shows a marked peak at
0.287 $\pm0.048$ c d$^{-1}$ (3.48 $\pm0.7$ d) where the error is based on the FWHM of the
peak. There are peaks of much lower significance in 2005.

Figure 3 shows the light curves folded on a double cycle of the planetary orbital period of 3.3125 d \citep{Lei03}
where zero phase corresponds to the planet being directly between the observer and the primary. 0.05 phase means are shown by solid points and the satellite orbital means by open circles (the errors are smaller than the points in all cases). In 2004, a shallow minimum $\sim$1 mmag near phase 0.8 is obvious, while in 2005, there are both shallow minima and also a maximum at phase 0.8 with peak to peak variations $\sim$1 mmag. Enhanced variability can also be seen within the 2004 minimum.

%______________________________________________ 
%______________________________________________ 

%______________________________________________ 
   \begin{figure}[htb]
   \centering
\includegraphics[height=0.48\textwidth, angle=270, bb=50 30 584 770]{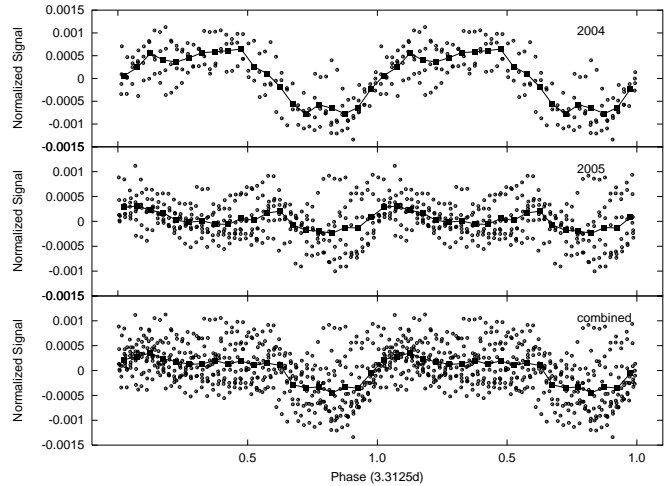}
\caption{The {\it MOST} light curves of $\tau$ Boo from 2004 and 2005 phased separately and together to two cycles of the planetary orbital period of 3.3125 d. The solid points are 0.05 phase
mean values and the open circles are satellite orbital means.
Errors are smaller than the sizes of the points in all cases. Zero phase corresponds to the planet lying directly between the observer and the primary.}
\label{FigPhase}
\end{figure}
%______________________________________________ 

\section{{\it MAD} Analysis}

To properly quantify the apparent activity as a function of phase, especially in 2005, we calculate the mean absolute deviation ({\it MAD}) signal. We use the following formalism to calculate the {\it MAD} signal.
The index of the individual phase bins [1 to 20] is $i$, and $n_{i}$ is the total number of data points in each bin where $k$ is the index of the signal values in each phase bin [1 to $n_{i}$].
The mean reference is the mean of the entire data-set and $MAD_{i}$ values for each phase bin  are calculated from: 
\begin{displaymath}
MAD_{i} = n_{i}^{-1}\sum_{k=1}^{n_{i}}|data_{k}-mean|
\end{displaymath}
We use a phase minimum of T$_{0}$=2452892.864  ($JD$-2451545 = 1347.864  d - the sub-planetary point),  and calculate the 
{\it MAD} signal using the {\it MOST} data binned every orbital period
(the large solid points of Figure 1). 

We calculated the {\it MAD} signal for a range of periods for the 2004, 2005, and 2004 \& 2005 data. 
For every {\it MAD} signal we searched for the most significant box-shaped increase in the {\it MAD} curve. To quantify the strength of such
a bump we used the 
edge effect box-fitting least squares (BLS) signal-to-noise statistic of \citet{Kov02}.
Alternative means of quantifying the strength of a feature in the {\it MAD} curve are expected to give similar results. 
The BLS signal-to-noise statistic for a range of periods for the 2004, 2005 and 2004 \& 2005 data is shown in Figure 4.
20000 period steps were used. The BLS algorithm was restricted to search for 
box-shaped signals with durations in phase from 0.25 to 0.45.
We tested the range in periods from 2.6 to 4.1 d reported by  \citet{Hen02} from Ca II H \& K observations. 

\begin{figure*}[th]
   \centering
\includegraphics[height=0.8\textwidth, angle=270]{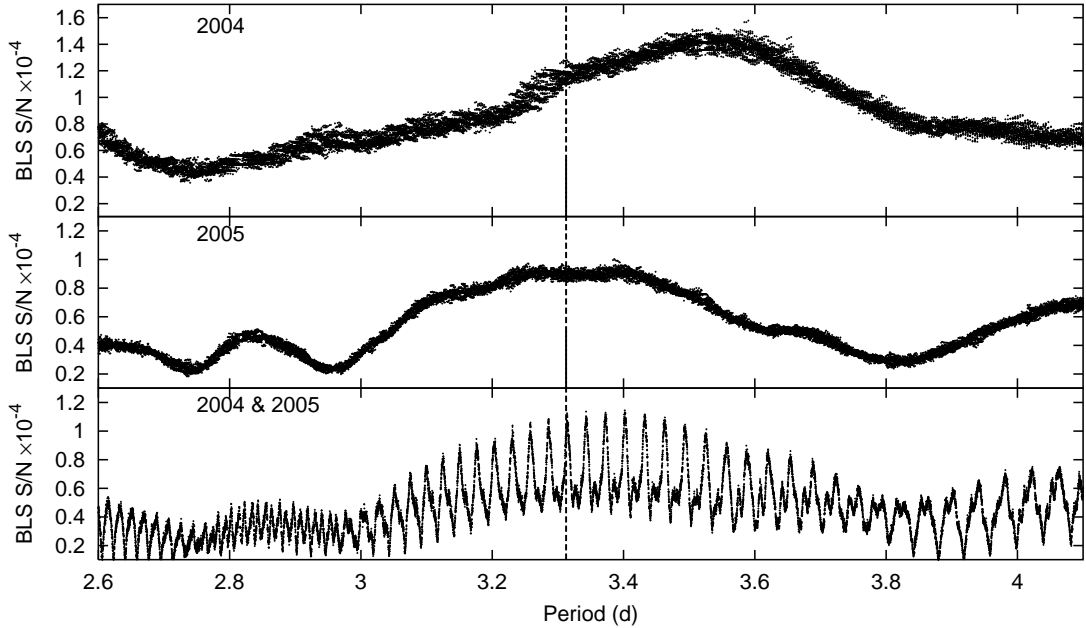}
\caption{ The {\it MAD} box-fitting least squares (BLS) signals calculated for a range of periods at the same vertical scale
for 2004 (top), 2005 (middle), and 2004 \& 2005 (bottom).
The vertical dashed line in each panel indicates the 3.3125 d orbital period of the planet, $\tau$ Boo b.
The apparent beating phenomenon displayed in the 2004 \& 2005 (bottom) panel is due to the one-year gap between our observations.}
\label{FigMADBryce}
\end{figure*}

As can be seen in Figure \ref{FigMADBryce}, the most significant {\it MAD} range in periods displayed in the 2005 light-curve is approximately
\PLowMADFive \ - \PHighMADFive  d (within the period range reported by \citealt{Cat07}).
The most significant {\it MAD} range in periods displayed in 2004 occurs
at longer periods: \PLowMADFour \ - \PHighMADFour \  d, which does not exclude the orbital period.
For the combined 2004 \& 2005 data the most significant
{\it MAD} range in periods is \PLowMADFourFive \ - \PHighMADFourFive \  d.
The apparent beating phenomenon apparent in the 2004 \& 2005 data 
is due to the one-year gap between our observations.
For the 2004 data, as the signal is strongly
sinusoidal, we argue that the Fourier Transform analysis discussed
above ($\S$\ref{SecFT})
is a much better indicator of the most significant period for the observed modulation,
rather than the {\it MAD} signal.

%______________________________________________ 
%______________________________________________ 
   \begin{figure*}[th]
   \centering
\includegraphics[height=0.8\textwidth, angle=270]{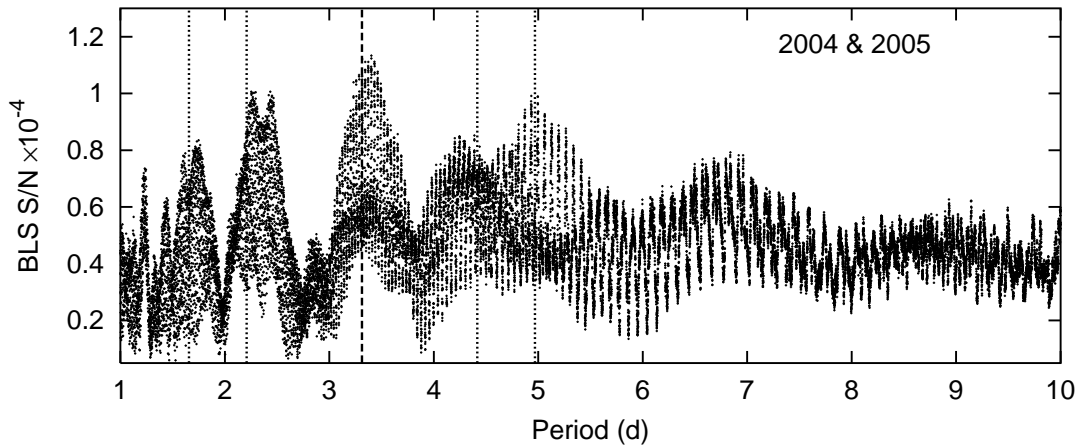}
\caption{The {\it MAD} box-fitting least squares (BLS) signals
calculated for a wide-range of periods for the combined 2004 \& 2005 data-sets.
The vertical dashed line indicates the 3.3125 d orbital period of the planet, $\tau$ Boo b,
while the vertical dotted lines indicate the 1/2 ($\sim$ 1.66 d), 2/3 ($\sim$ 2.21 d), 4/3 ($\sim$ 4.42 d),
and 3/2 ($\sim$ 4.97 d) harmonics and subharmonics of this orbital period.
The apparent beating phenomenon is due to the one-year gap between our observations.
%The section from 2.6 - 4.1 d is identical to the bottom panel of Figure $\ref{FigMADBryce}$, except with coarser frequency resolution.
}
\label{FigMADExpanded}
\end{figure*}
%______________________________________________ 

We repeated the above {\it MAD} analysis over a wider range of periods so
as to rule out tidal-interactions, and longer term variability. For this extended {\it MAD} analysis
30000 period steps were tested with periods from 1.0 to 10.0 d. 
These results are displayed in Figure \ref{FigMADExpanded}.
As can be seen in this figure the most significant {\it MAD} signal is near that of the planetary period.
The apparent beating phenomenon, due to the one-year gap between our observations, is again obvious.
The other peaks in this figure that approach the significance of the main peak are all near
harmonics and subharmonics of the main peak, and for this reason are not believed
to be significant.

In Figure \ref{FigMAD} the {\it MAD} signals for $\tau$ Boo are plotted separately and together from 2004 and 2005 phased to the planetary orbital period of 3.3125 d.
This shows the increased activity in 2005 near $\phi=0.8$. There is a similar but less well differentiated enhancement in 2004 with the combined data showing the most obvious effect.

\section{Synchronism of the 2004 and 2005 Light curves.}
\label{synchro}

The independent phasing to the planetary orbital period of the 2004 dark spot and the variable region in 2005 is insufficiently precise to make a convincing case for synchronism with the planetary period. The 'spot' in 2004 occurs at approximately 0.875 $\pm$0.05 in phase, while the 2005
variable region (determined from Figure \ref{FigMAD}) is at phase 0.825 $\pm$0.05. But,
there were 406.6 days between the first {\it MOST} observation of $\tau$~Boo A
in 2004 and the final observation in 2005,
equivalent to nearly 123 planetary revolutions. If the spot and the variable region from 2004 and 2005 are the same then they are synchronised with the orbital period of the planet ($P$ = 3.31245 d) to better than 0.04\% of the planetary period.

\section{Ca II K line activity}
\label{CaIIK}

As part of the search for chromospheric activity synchronised to planetary orbital periods in 51 Peg systems,
\citet{Shk05} observed Ca II H \& K reversals on fifteen  nights for $\tau$~Boo over four semesters in 2001, 2002 (2 semesters)
and 2003. The K line residuals from an overall mean are plotted as a function of orbital phase in Figure 7 of \citet{Shk05}.
Like \citet{Hen00}, they found the general level of chromospheric activity to be low with no obvious enhancement synchronised to
the orbital period. 

Here, we look instead for enhanced {\it variability} which might mimic that seen in white light by {\it MOST}. In Figure \ref{FigCaHK} {\it MAD}
values for the residual K line emission are shown averaged over two phase bins: one during the photometrically determined active period,
phases 0.6 to 1.1, and the other outside of it, phases 0.1 to 0.6. The error bars span twice the average measurement error within each bin.
Despite the average intensity level for the K line differing from year to year, no attempt was made to align the average annual values.
There appears to be a significant increase in {\it variability} in the 0.8 phase centered bin relative to the other.

% If one assumes that the variable region detected by {\it MOST} on  $\tau$~Boo has been long lived then the Ca II K line reversal intensities from the three years prior
% to the {\it MOST} photometry give support to the conclusion that there was a significantly enhanced chromospheric and white light variability stimulated in the equatorial region of the star some 68$^{\circ}$ ahead of the sub-planetary point for four years or some 440 orbital revolutions. 

If the enhanced variability of the residual Ca II K line reversal intensities in 
the 0.8 centered phase bin 
in fact results from the same 
variable region detected by {\it MOST} on $\tau$~Boo, 
than this result supports the conclusion
that there was significantly enhanced chromospheric and white light variability some 68$^{\circ}$ ahead of the
sub-planetary point for four years or some 440 orbital revolutions. 

\section{Conclusions}
\label{Conclusions}
The {\it MOST} light curves presented in this paper from 2004 and 2005 are the `most' continuous and accurate ever obtained for $\tau$ Boo and offer a unique opportunity to look for the subtle  photometric effects of star-planet interaction in this 51 Peg-type system.

\begin{figure}[htb]
   \centering
\includegraphics[height=0.5\textwidth, angle=270, bb=50 50 584 770]{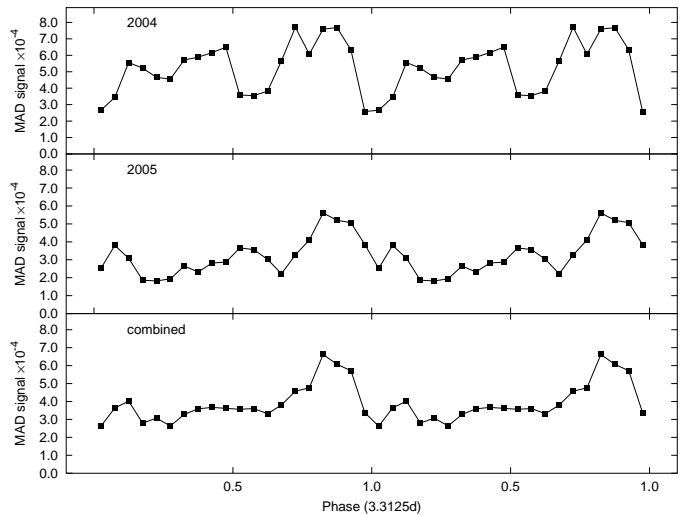}
\caption{The {\it MOST} {\it MAD} signals for $\tau$ Boo from 2004 and 2005 phased separately and together to two cycles of the planetary orbital period of 3.3125 d. Error bars are smaller than the points in all cases. Zero phase corresponds to the planet lying directly between the observer and the primary.}
\label{FigMAD}
\end{figure}

To the eye, the 2004 curve (Figure 1) shows a persistent, periodic dark `spot' causing $\sim$1 mmag dimming, roughly synchronised to the planetary orbital period. A similar intermittent `spot' is seen in the longer run of 2005.  We have tried to quantify the effects in this paper.  A Fourier transform of the 2004 light curve (Figure 2) gives a period of 3.5 $\pm$0.7 d but no signal of similar strength in 2005. However, the obvious activity in 2005 shows a well defined maximum in the {\it MAD} values (half width $\sim$0.15 d) centered on the planetary orbital frequency of 3.3125 d (Figure 4).
The spot in 2004 and the variable region defined by the maximum {\it MAD} signal
in 2005 both precede the sub-planetary longitude on the primary by $\sim$68$^{\circ}$ ($\phi=0.8$) when the light curves are phased to the planetary orbital period (Figures 3 \& \ref{FigMAD}).

\begin{figure}[htb]
   \centering
\includegraphics[width=0.48\textwidth]{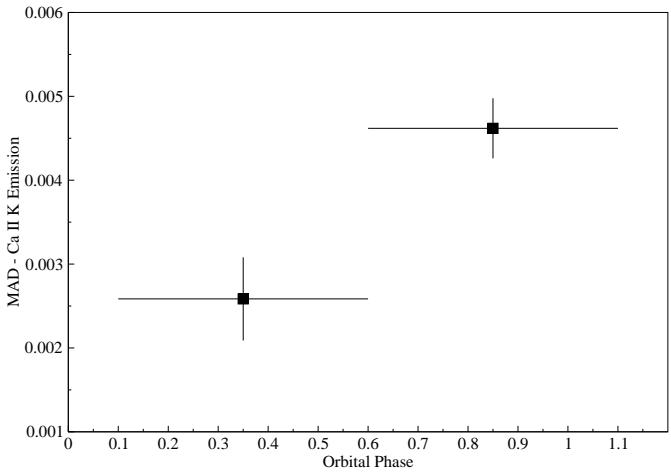}
\caption{Average mean absolute deviations ({\it MAD}) for the combined 2001, 2002 (2 data sets), and 2003 integrated Ca II K line flux residuals from \citet{Shk05} (see their Figure 7) phased to the 3.3125 day orbital period of $\tau$ Boo b. One phase bin corresponds to the photometrically active period found by {\it MOST}, $\phi$ = 0.6 -- 1.1, and the other to the less active period, $\phi$ = 0.1 -- 0.6. The error bars are double the average measurement error within each bin.}
\label{FigCaHK}
\end{figure}

If the variable region has in fact persisted over the 123 planetary orbits covered by the 2004 and 2005 light curves then the active spot would be synchronized to the planetary orbital period to within \PlanetSynch \ d (Figure 4). The $\tau$ Boo A Ca II K line in 2001, 2002 and 2003 also displayed enhanced K-line variability centered on $\phi=0.8$ (Figure \ref{FigCaHK}) which potentially extends the detection to some 440 orbital periods suggesting that this is indeed a really long lived feature.

Because {\it MOST} detected but a single `spot' with a unique period it is not possible to say anything about differential rotation of the primary from the photometry but, both \citet{Hen00} and \citet{Cat07} have presented good evidence for differential rotation from Ca II H \& K photometry and magnetic field variations. The presence of a global stellar magnetic field implies regeneration by dynamo action for which the engine is generally assumed to be differential rotation (see for example \citealt{Oss03}). Given that spots like those on the Sun are expected to gradually migrate and appear at different latitudes, our evidence that the variable region has persisted for more than 4 years 68$^{\circ}$ in advance of sub-planet point suggests that it owes its existence and rotation period to the planet rather than stellar rotation.

The existence of only a single `spot' significantly in advance of the sub-planetary point effectively rules out the possibility that
the activity is tidally induced and makes magnetic interaction the most likely candidate.
The rapidity of the variability on stellar rotational time scales makes interpretation in terms of conventional stellar spot groups unlikely, but the weakness of the signals presented here does not  allow us to totally rule out that possibility.
Regardless, we feel the most plausible explanation for the observed modulation of the light curve
is an active region magnetically induced on the star by the planetary companion.

%______________________________________________ 
%______________________________________________ 
%______________________________________________ 
%______________________________________________ 

Based on the cyclical dwell time or visibility of the active region in 2004 and the inclination of the rotation axis of 40$^{\circ}$ suggested by \citet{Cat07}, the active region appears to lie closer to a latitude of +30$^{\circ}$ than to the equator. If true, this would lend support to the contention of \citet{Cat07} that the active region has developed at a latitude in synchronous rotation with the planet. For the moment, this remains speculation until we have a better understanding of the excitation mechanism. It will be very important to continue attempts to map the magnetic field on the primary and its variations.

Taken together, we feel that the {\it MOST} photometry and the monitoring of Ca II H \& K  over four years make a credible case for variability 
resulting from an active region induced magnetically on  the surface of $\tau$ Boo A by its planetary companion.

\begin{acknowledgements}
The Natural Sciences and Engineering Research Council of Canada supports the research of B.C., D.B.G., J.M.M., A.F.J.M., S.M.R.. Additional support for A.F.J.M. comes from FQRNT (Qu\'ebec). R.K. is supported by the Canadian Space Agency. W.W.W. is supported by the Austrian Space Agency and the Austrian Science Fund (P17580). DH is supported by the Austrian Fonds zur F\"orderung der wissenschaftlichen Forschung, project number P17580-N02
\end{acknowledgements}

\end{document}